\input{aipcheck}

\documentclass[
    ,final            
  ]
  {aipproc}

\layoutstyle{6x9}

\begin{document}

\title{Black-hole binaries: life begins at 40 keV}

\classification{97.10.Gz -- 97.60.Lf -- 97.80Jp}
\keywords      {X-ray: binaries -- accretion: accretion disks -- black hole: physics}

\author{Tomaso M. Belloni}{
  address={INAF-- Osservatorio Astronomico di Brera, Via E. Bianchi 46, I-23807 Merate, Italy}
}

\author{Sara Motta}{
  address={Universit\`a dell'Insubria, Como, Via Valleggio 11, I-22100 Como, Italy \& INAF-- Osservatorio Astronomico di Brera, Via E. Bianchi 46, I-23807 Merate, Italy}
}

\begin{abstract}
In the study of black-hole transients, an important problem that still needs to be answered is how the high-energy part of the spectrum evolves from the low-hard to the high-soft state, given that they have very different properties. Recent results obtained with RXTE and INTEGRAL have given inconsistent results. With RXTE, we have found that the high-energy cutoff in GX 339--4 during the transition first decreases (during the low-hard state), then increases again across the Hard-Intermediate state, to become unmeasurable in the soft states (possibly because of statistical limitations). We show Simbol-X will be able to determine the spectral shape with superb accuracy. As the high-energy part of the spectrum is relatively less known than the one below 20 keV, Simbol-X will provide important results that will help out understanding of the extreme physical conditions in the vicinity of a stellar-mass black hole.
\end{abstract}

\maketitle


\section{Introduction}

Black-hole transients, during their outbursts, undergo strong changes in both spectral and timing parameters. These variations can be classified through the introduction of a small number of source states (see \cite{belloni2005,hombel2005,belloni2009}). The most important transition is that from the Hard-Intermediate State (HIMS) to the Soft-Intermediate State (SIMS), which takes place at high luminosity. Part of the full transition from the hard to the soft state, the HIMS-SIMS transition has been associated to the ejection of powerful transient relativistic jets (see \cite{fbg04,fhb09}). Through this transition, spectral parameters below 20 keV change little, while strong an fast changes are observed in the properties of fast timing. 

In the hard state (LHS), the energy spectrum shows a clear high-energy cutoff around 100 keV, while in the soft state (HSS) a cutoff was not observed up to $\sim$1 MeV (Grove et al. 1998). These spectra are interpreted as the result of Comptonization from a combination of thermal and non-thermal electrons (see e.g. \cite{Ibragimov}). Different theoretical models differ in the details of the physical location and combination of these components.

A promising way to discriminate between models is to observe in detail the transition from the hard to the soft state,  through which the energy spectrum changes between two extremes. This will also allow to determine what physical changes take place in the accretion flow across the HIMS-SIMS transition, closely related to the jet ejection. However, the full hard to soft transition can be very short, of the order of a few days, and the HIMS-SIMS transition is faster than one day. It is therefore difficult to obtain the time coverage needed for this study.

\section{The riddle of the hard component}

Outbursts of black-hole transients are observed several times by RossiXTE. In 2004, an INTEGRAL campaign to observe GX 339--4 was started during a series of RossiXTE observations with the aim of detecting a transition. The transition took place during the INTEGRAL coverage (see Fig. 1, left panel) and high-quality combined spectra could be obtained \cite{belloni2006}. The transition was established from RossiXTE timing data and the energy spectra showed a clear evolution that took place on a time scale below 12 hours. At the end of the HIMS, a high-energy cutoff was present at around 70 keV, while after the transition no cutoff was detected with a lower limit of 100 keV \cite{belloni2006}.
In 2007, GX 339--4 underwent a new outburst. A similar combined RossiXTE-INTEGRAL campaign covered only a secondary transition between HIMS and SIMS (\cite{delsanto2009}, Fig.1 left panel). During this transition, the high-energy cutoff was observed to {\it decrease}, opposite to what observed in 2004. It is clear from Fig. 1 that the coverage of these two campaigns was good but still not optimal.

\begin{figure}
\begin{tabular}{cc}
  \includegraphics[height=.27\textheight]{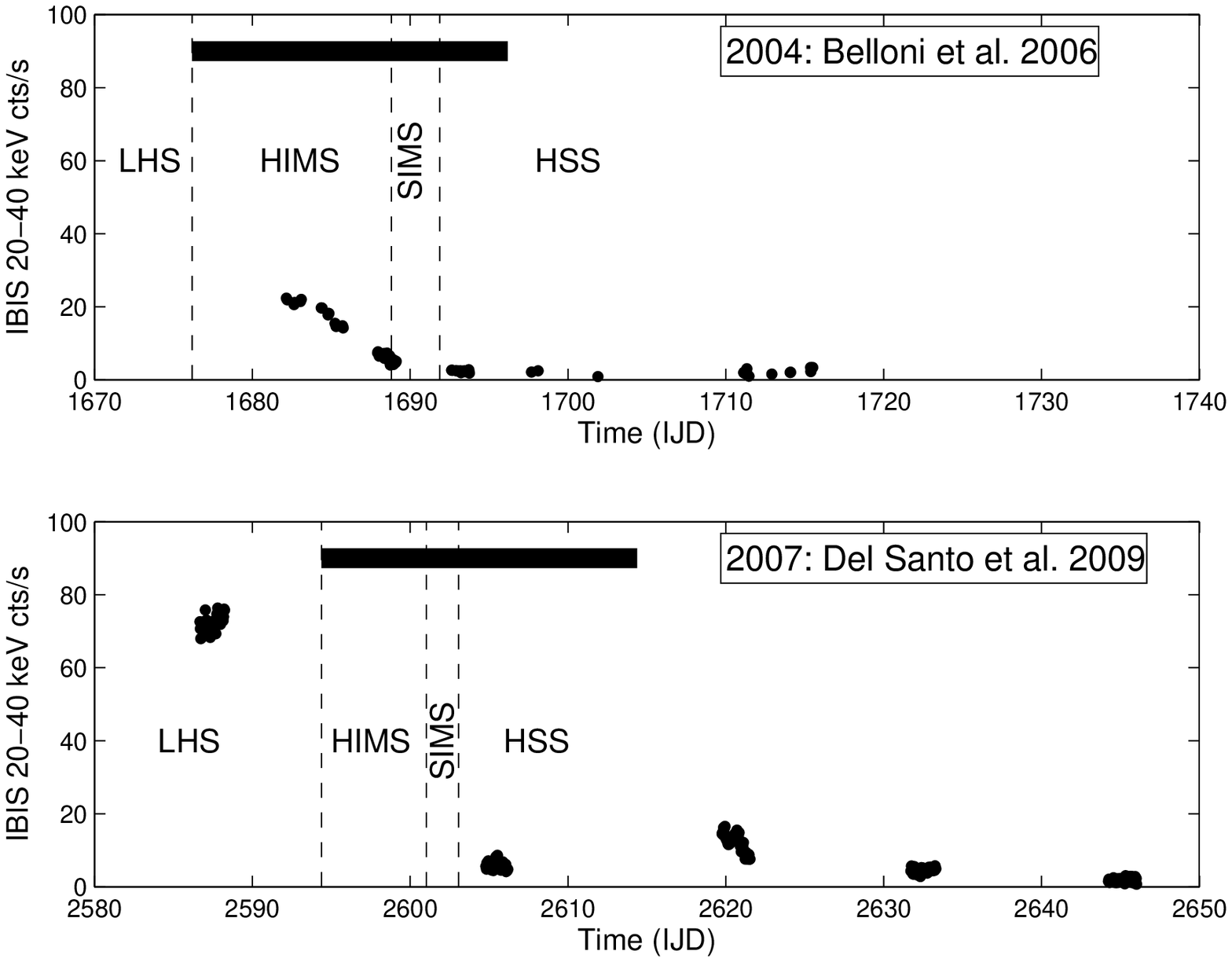} &
  \includegraphics[height=.27\textheight]{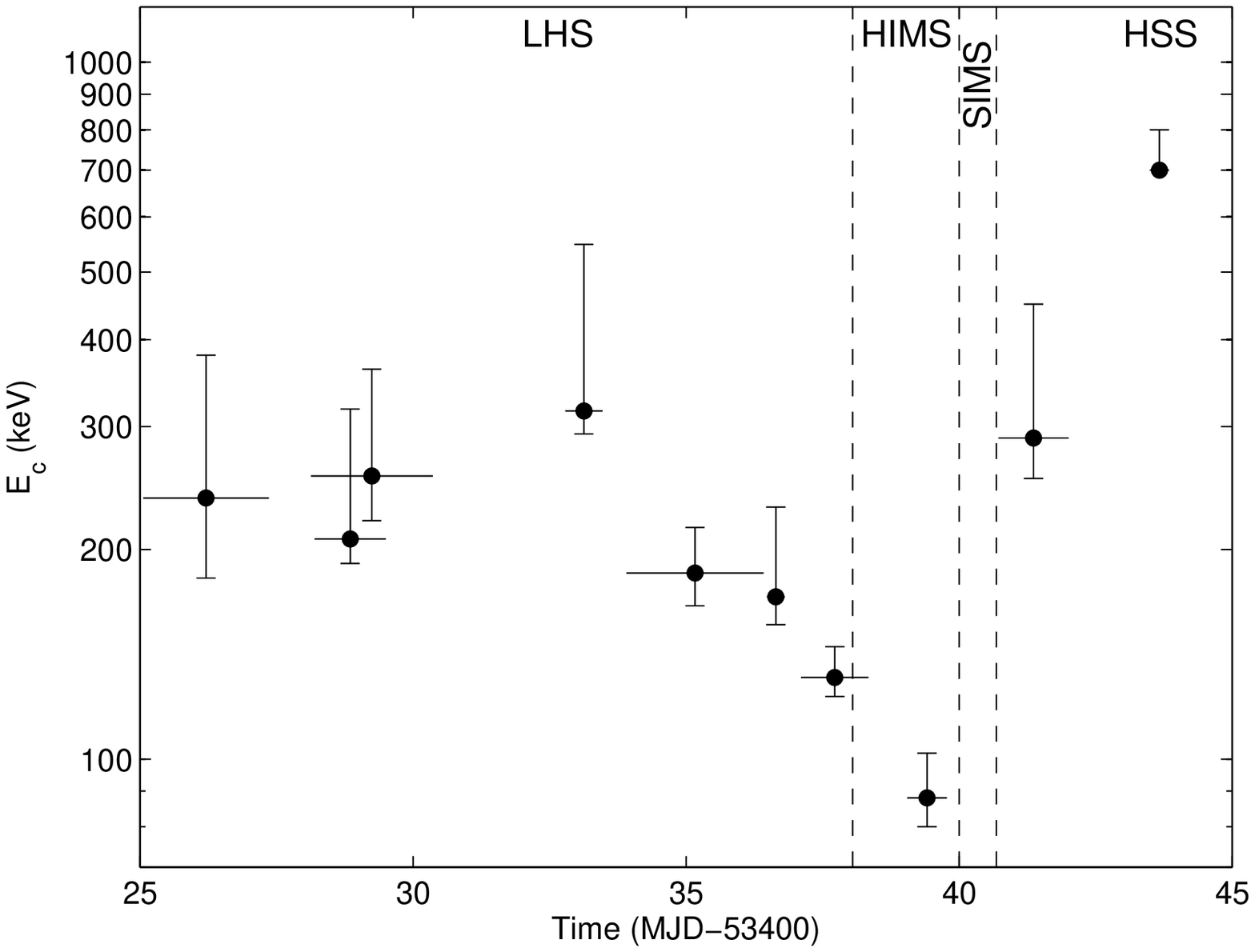} \\
\end{tabular}
    \caption{Left panel: INTEGRAL/IBIS light curves for the 2004 (top) and 2007 (bottom) outbursts of GX 339--4. The states as measured by RossiXTE are marked. The horizontal black bars indicate the period with RossiXTE coverage.
    Right panel: time evolution of the high-energy cutoff for the 2005 outburst of GRO J1655--40 (from the data of \cite{joinet2008}), with RossiXTE-determined states.}
\end{figure}

An indication of the possible solution of this inconsistency can be found in the 
INTEGRAL-RXTE-Swift data of the 2005 outburst of the bright transient GRO J1655--40 \cite{joinet2008}. From their table, it is possible to reconstruct the time evolution of the high-energy cutoff, shown in the right panel of Fig. 1. The cutoff decreases through the hard state and the HIMS, to increase again in SIMS and soft state. Here the coverage is good, but unfortunately the HIMS was particularly short for this outburst.

\section{New results}

The 2007 outburst of GX 339--4, in addition to the secondary transition reported by \cite{delsanto2009} also underwent a major transition from hard to soft earlier in the ourburst. This transition lasted around ten days and its coverage with RossiXTE was very good. The full analysis of the RossiXTE data is presented in a forthcoming paper \cite{motta2009}. Particularly interesting is the behavior of the high-energy cutoff as a function of time across the transitions, LHS--HIMS--SIMS--HSS, which can be seen in Fig. 2. The figure shows clearly a cutoff that decreases monotonically from 130 keV to 60 keV as the source brightens (and softens) in the LHS, than it increases back to $\sim$100 keV during the HIMS, when the energy spectrum softens also because of the appearance of the thermal disk. In the SIMS and the HSS, the measured cutoff is as large as 160 keV. Clearly, a non-monotonic evolution can explain the inconsistent results obtained earlier for the same source.

\begin{figure}
  \includegraphics[height=.30\textheight]{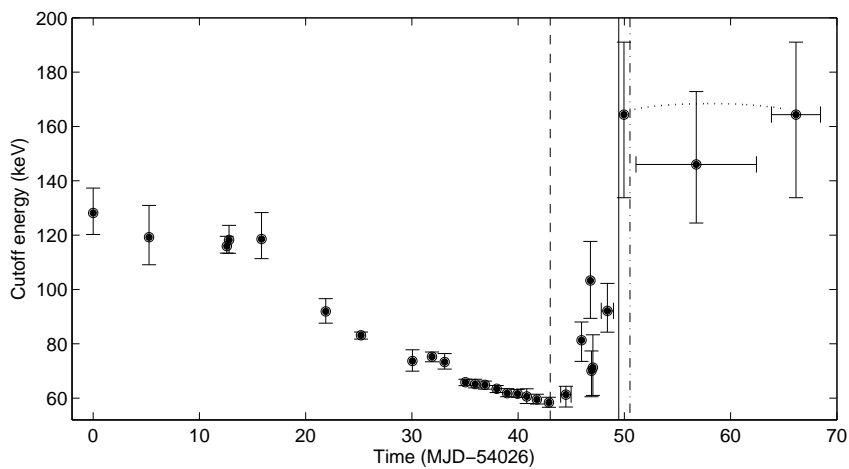}
    \caption{Time evolution of the high-energy cutoff in GX~339--4 during its 2007 outburst (from \cite{motta2009}). The three vertical lines mark the LHS-HIMS (dashed), HIMS-SIMS (filled) and SIMS-HSS (dot-dash) transitions. The dotted ``slur'' indicates that those two points belong to the same spectral fit.
    }
\end{figure}

\section{Discussion: prospects for Simbol-X}

The problem we have is to discover how to go from a 100 keV cutoff to a very-high or no cutoff at all. The right panel of Fig. 1 and Fig. 2 show the first attempt to give a global answer to this question. Contrary to what could be expected, the evolution is not from 100 keV up, but is non-monotonic. First, during the LHS, the cutoff decreases by a factor of two, indicating a significant cooling of the emitting electrons. After the transition to the HIMS, i.e as the transition starts, this evolution is reversed at the cutoff increases, to become difficult to detect in the HSS. 

The results shown above make it clear that the high-energy cutoff in black-hole binaries can also be considerably lower than the canonical 100 keV, in particular during crucial transitions that constitute the most interesting and crucial part of the outbursts. 
In order to test the capabilities of Simbol-X, we have used the current detector responses available for simulations and reproduced a set of ten spectra as observed by RossiXTE from GX 339--4 in 2007 (see Fig. 2). Simple spectral fitting allowed us to recover the value of the high-energy cutoff with high precision all over the interval, as can be seen from the tiny size of the error bars, most of which are within the symbols.
While it is not expected that Simbol-X performs a dense campaign such as those of RossiXTE on any transient, it is clear that a few observations in strategically planned parts of the outburst will be able to yield superb spectra which can then be used to test specific physical models for the hard emission.
The high-energy part of the spectrum is very important and is relatively less known than the one below 20 keV, for which many observations at high statistical level are available, Simbol-X has the potential to provide important results that will help out understanding of the extreme physical conditions in the vicinity of a stellar-mass black hole.

\begin{figure}
  \includegraphics[height=.27\textheight]{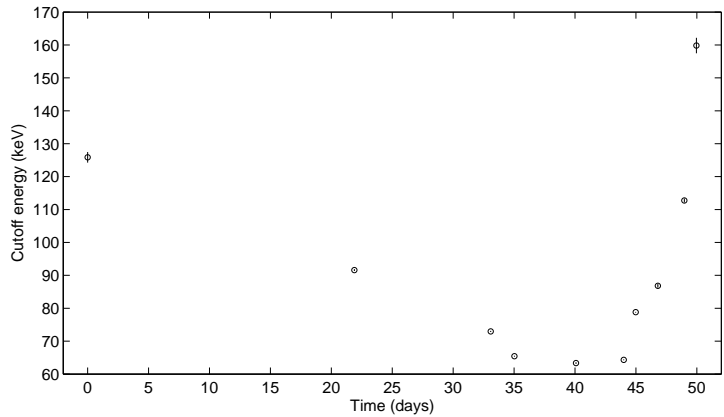}
    \caption{Time evolution of the high-energy cutoff in GX~339--4 during its 2007 outburst (from \cite{motta2009}), simulated through a few observations with Simbol-X and reconstructed through spectral fits. The exposure time for each observation is 100 ks.
    }
\end{figure}


\begin{theacknowledgments}
We acknowledge support from ASI via contract ASI/INAF I/023/05/0.
\end{theacknowledgments}


 \end{document}